\documentstyle[prl,aps,epsf,here]{revtex}
\twocolumn

\def\sinh{\hbox{sinh}}
\def\ctgh{\hbox{ctgh}}
\def\tgh{\hbox{tanh}}
\def\hom{\hbar\omega}
\def\eq#1{(\ref{#1})}
\def\fig#1{Fig.~\ref{#1}}
\newcommand{\etal}{{\em et al.{}}}
\newcommand{\be}{\begin{equation}}
\newcommand{\ee}{\end{equation}}
\newcommand{\bea}{\begin{eqnarray}}
\newcommand{\eea}{\end{eqnarray}}
\newcommand{\Figurebb}[9]{
  \begin{figure}[H]\begin{center}
  \leavevmode
  \epsfysize=#7cm
  \epsfbox[#2 #3 #4 #5]{#6}
  \par
  \parbox{#8cm}{
  \caption[figure]{\renewcommand{\baselinestretch}{0.8} \small
                   \hspace{-0.3truecm}#9}\label{#1}}
  \end{center}
  \end{figure}
}

\begin{document}

\draft
\title{Simple Analytical Particle and Kinetic Energy Densities for a
       Dilute Fermionic Gas\\ in a $d$-Dimensional Harmonic Trap}
\author{Matthias Brack}
\address{Institut f\"ur Theoretische Physik, Universit\"at Regensburg,
         D-93040 Regensburg, Germany}
\author{Brandon P. van Zyl}
\address{Institute for Microstructural Sciences, National Research Council
         of Canada, Ottawa, Ontario, K1A OR6}
\date{\today}
\maketitle

\widetext
\vspace*{-6.2cm}
\begin{flushright}
{\bf TPR-00-17}
\end{flushright}
\vspace*{5.2cm}
\narrowtext

\vspace*{-0.73cm}

\begin{abstract}
We derive simple analytical expressions for the particle density
$\rho(r)$ and the kinetic energy density $\tau(r)$ for a system of
noninteracting fermions in a $d$-dimensional isotropic harmonic
oscillator potential. We test the Thomas-Fermi (TF, or local-density)
approximation for the functional relation $\tau[\rho]$ using the exact
$\rho(r)$ and show that it locally reproduces the exact kinetic energy
density $\tau(r)$, {\it including the shell oscillations,} surprisingly
well everywhere except near the classical turning point. For the
special case of two dimensions (2D), we obtain the unexpected
analytical result that the integral of $\tau_{TF}[\rho(r)]$ yields the
{\it exact} total kinetic energy.
\end{abstract}


\vspace*{-0.28cm}

The evaporative cooling of dilute (i.e., almost noninteracting)
fermionic gases has recently been achieved by Jin and DeMarco at JILA
in Colorado \cite{jin}. This spectacular experimental milestone has
stimulated an enormous effort to explore and understand the properties
of these new quantum systems, which can be viewed as the quantum
analogue of the Bose-Einstein condensation (BEC) recently observed in
ultracold trapped Bose gases. While it is true that the theory of
homogeneous dilute Fermi systems is fairly well developed, addressing
new experiments probing strongly inhomogeneous systems in regimes far
from equilibrium, represents a much greater challenge to theory. To
this end, Vignolo \etal\ \cite{vign} have recently used a Green's
function method to compute the particle and kinetic energy densities
for a system of noninteracting fermions in a one-dimensional (1D)
harmonic oscillator potential.  In principle such a (quasi)-1D system
can be achieved experimentally using state-of-the-art magnetic
confinement techniques \cite{jin}. Owing to the enhanced shell
structure found in 1D, Vignolo {\etal} suggest that these quantum
oscillations may be accessible to observation in magnetically trapped
gases of fermionic alkali atoms. We give here much simpler {\em
analytical} results for the more general case of a $d$-dimensional
harmonic potential and use them to test their Thomas-Fermi (TF)
functional relation.

The work presented in this paper is also applicable to a 2D electron
gas confined to so-called quantum dots \cite{dots}. The external
confinement potential of these dots is in many cases essentially
harmonic. Bhaduri \etal\ \cite{frac} have shown that the inclusion of
a short-range two-body interaction may be included via fractional 
statistics, provided that one uses the TF relation $\tau_{TF}[\rho]$ 
relevant for 2D.

{\it The method. ---} We start from a system of noninter\-acting
fermions described by the time-independent Schr\"odinger equation
\be
{\hat H}\phi_i({\bf r}) = \left[ {\hat T} + V({\bf r})\right]
                          \phi_i({\bf r})
                        = \epsilon_i\phi_i({\bf r})\,,
\ee
where $V(\bf r)$ is a local potential to be specified later. The
single-particle density matrix can be obtained by an inverse Laplace
transform of the Bloch density matrix:
\be
\rho({\bf r,r'}) = 2\!\! \sum_{\epsilon_i<E_F}
                   \phi_i^*({\bf r'}) \phi_i({\bf r})
                 = {\cal L}_{E_F}^{-1}\left[\frac{2}{\beta}\,
                   C({\bf r,r'};\beta)\right]\!,
\ee
where the latter is defined by
\be
C({\bf r,r'};\beta) =  \sum_{\hbox{\footnotesize all }i}
                       \phi_i^*({\bf r'})
                       \phi_i({\bf r})\,\exp\{-\beta\epsilon_i\}.
\ee
$E_F$ is the Fermi energy, the factor 2 accounts for spin.
We now use center-of-mass and relative coordinates:
\be
{\bf q} = \frac12\,({\bf r}+{\bf r'}), \quad {\bf s}
        = {\bf r}-{\bf r'},
\ee
so that the local density is $\rho({\bf q})=\rho({\bf q,s})|_{{\bf
s}=0}$. For the kinetic energy density, we investigate two expressions
\cite{tau}:
\bea
\tau({\bf q}) & = & - \frac{\hbar^2}{2m}\, 2\!\! \sum_{\epsilon_i<E_F}
             \phi_i^*({\bf q})\nabla^2 \phi_i({\bf q})\,,\label{tau}\\
\tau_1({\bf q}) & = & \frac{\hbar^2}{2m}\, 2\!\! \sum_{\epsilon_i<E_F}
              |\nabla\phi_i({\bf q})|^2\,.\label{tau1}
\eea
In the presence of time-reversal symmetry they are simply related by
\be
\tau({\bf q}) = \tau_1({\bf q}) + \frac12\, \frac{\hbar^2}{2m}\,
                                  \nabla^2\rho({\bf q})\,.
\ee
A convenient quantity is their mean,
\be
\xi({\bf q}) = \frac12\, [\tau({\bf q})+\tau_1({\bf q})]\,,
\ee
which is obtained from the density matrix by
\be
\xi({\bf q}) = -\frac{\hbar^2}{2m}
                \left[\nabla^2_s \rho({\bf q,s})\right]_{s=0},
\label{xi}
\ee
where $\nabla_s$ is the gradient with respect to the variable $\bf s$.
Note that all three quantities $\tau({\bf q})$, $\tau_1({\bf q})$, and
$\xi({\bf q})$ integrate to the same exact kinetic energy.

We now specialize to an isotropic harmonic oscillator potential in $d$
dimensions:
\be
V(r)= \frac{m}{2}\,\omega^2 r^2,
\ee
where $r=\sqrt{x_1^2+\dots+x_d^2}$ is the radial variable. The exact
Bloch density matrix for this system is given by \cite{sond}
\bea
& & C({\bf r,r'};\beta)= C(q,s;\beta)=
       \left(\frac{m\omega}{2\pi\hbar}\right)^{\!d/2}
       \frac{1}{\sinh^{d/2}(\beta\hom)}
       \times \nonumber \\
& &  \quad\exp\left\{-\frac{m\omega}{\hbar}
     \left[q^2\tgh\!\left(\!\frac{\beta\hom}{2}\!\right)\! +\!
     \frac{s^2}{4}\ctgh\!\left(\!\frac{\beta\hom}{2}\!\right)
     \right]\right\}.
\label{cho}
\eea
To get the particle and kinetic energy densities, we need to perform
inverse Laplace transforms of the above function and its derivatives at
$s=0$. For the first exponential factor in \eq{cho}, we employ the
following relation which can be derived from Ref.\ \cite{grad}
\bea
& & \exp\left\{-x\,\tgh(\beta/2)\right\} \nonumber \\
& & \qquad = \sum_{n=0}^\infty (-1)^n
           L_n(2x)\,e^{-x} \left\{e^{-n\beta}+e^{-(n+1)\beta}\right\}.
\label{tghid}
\eea
This relation holds if $|z|=|e^{-\beta}|<1$, which is fulfilled since
the contour of the inverse Laplace transform integral in the complex
$\beta$ plane goes along $\beta=it+c$ for $t\in (-\infty,\infty)$ with
$c>0$. Further analytical progress depends on the dimensionality $d$.

{\it The case $d=2$. ---}
Here we can directly
use the following exact Laplace inverse \cite{abro}
\be
{\cal L}_\lambda^{-1}\left[\frac{e^{-n\beta}}{\beta\,\sinh(\beta)}\right]
=2\sum_{k=0}^\infty \Theta[\lambda-(2k+1)-n]\,.
\label{isinh}
\ee
When filling $M+1$ oscillator shells, the Fermi energy is
\be
E_F = \hom\,(M+1+\delta)\,,
\ee
with $\delta$ being an infinitesimally small positive number.
Combining Eqs.\ \eq{tghid}, \eq{isinh} and carefully evaluating the
sums over the step functions, we get for the density
\be
\rho(q) = 2 \left(\frac{m\omega}{\pi\hbar}\right) \sum_{\mu=0}^M
          (M\!-\!\mu\!+\!1)(-1)^\mu L_\mu(2x)\, e^{-x},
\label{den}
\ee
where $x=(m\omega/\hbar)\,q^2$.
The kinetic energy density \eq{xi} is given, after some suitable
manipulations of hyperbolic functions, by the following Laplace inverse:
\bea
& & \xi(q) = \hom \left(\frac{m\omega}{\pi\hbar}\right) \times \nonumber\\
       &   & \quad {\cal L}_{E_F}^{-1}\left[\frac{1}{\beta}
                   \frac{1}{2\,\sinh^2(\beta\hom/2)}
                   \exp\left\{-x\tgh(\beta\hom/2)\right\}\right].
\eea
Removing one inverse $\sinh$ factor by the identity
\be
\frac{1}{2\,\sinh(\beta\hom/2)} = \sum_{m=0}^\infty e^{-(m+1/2)\beta\hom}\,,
\ee
and proceeding as above, we get the final expression for the kinetic
energy density:
\be
\xi(q) = \hom \left(\frac{m\omega}{\pi\hbar}\right)\sum_{\mu=0}^M
          (M\!-\!\mu\!+\!1)^2(-1)^\mu L_\mu(2x)\, e^{-x}.
\label{kin}
\ee
The integrals $d^2q$ of the densities (\ref{den}), (\ref{kin}) are readily
evaluated using
\be
\int_0^\infty \!L_n(2x)\, e^{-x}\,dx = (-1)^n,
\ee
and yield the correct results for the number $N$ of particles in $M+1$
filled shells
\be
\int\! \rho(q)\,d^2q = 2\sum_{\mu=0}^M (\mu\!+\!1) = M^2\!+\!3M\!+\!2 = N(M)\,,
\ee
and for their exact kinetic energy $E_{kin}(M)$
\bea
& & \int \xi(q)\,d^2q = \hom \sum_{\mu=0}^M (\mu+1)^2 \nonumber\\
                & & \quad = \frac16\, \hom
                            \left(2M^3+9M^2+13M+6\right)
                          = E_{kin}(M)\,.
\eea

Next we investigate the Thomas-Fermi (TF) relation between $\tau$ (or
$\tau_1$ or $\xi$) and $\rho$, which in 2D is
\be
\tau_{TF}[\rho] = \frac{\hbar^2}{2m}\, \pi \rho^2,
\label{tau2ofrho}
\ee
see also Eq.\ \eq{tauofrho} below.
Inserting Eq.\ (\ref{den}) into the right-hand side above and
integrating, using the orthonormality of the Laguerre polynomials,
we find
\be
\int \tau_{TF}[\rho(q)]\,d^2q = \hom \sum_{\mu=0}^M (\mu+1)^2 = E_{kin}(M)\,.
\label{ekinex}
\ee
This means that the simple TF functional -- {\it without gra\-dient
corrections} \cite{weiz} -- using the exact density $\rho(q)$
yields the exact quantum-mechanical kinetic energy, which is highly
nontrivial and unexpected. The local behavior of $\tau_{TF}[\rho(q)]$
will be examined numerically below.

{\it The cases $d=1$ and $d \ge 3$. ---}
For $d=1$, we have a square root of the $\sinh$ factor in the
denominator of \eq{cho}. To handle it, we use the expansion
\bea
\sinh^{1/2}(s) & = & \frac{1}{\sqrt{2}}\left(e^s-e^{-s}\right)^{1/2}
                 =\frac{1}{\sqrt{2}}e^{s/2}\sqrt{1-e^{-2s}} \nonumber\\
               & = & \frac{1}{\sqrt{2}}e^{s/2}\left\{1-\sum_{m=1}^\infty
               \frac{(2m-3)!!}{(2m)!!}e^{-2ms}\right\},
\eea
which converges for Re $s>0$, and include it as a factor on top of the
$d=2$ case. For $d=3$, we have to include its inverse, which has
the expansion
\bea
& & \sinh^{-1/2}(s) = \sqrt{2}e^{-s/2}\left(1-e^{-2s}\right)^{-1/2} \nonumber\\
& & \qquad          = \sqrt{2}e^{-s/2}\left\{1+\sum_{m=1}^\infty
                      \frac{(2m-1)!!}{(2m)!!}e^{-2ms}\right\}.
\eea
For $d=4$, we need
\be
\sinh^{-1}(s) = 2\left(e^s-e^{-s}\right)^{-1}
              = 2\sum_{m=0}^\infty e^{-(2m+1)s}\,,
\ee
and so on. Using $E_F=\hom\,(M+d/2+\delta)$ and proceeding as above
for the Laplace inversions, we obtain the general expressions
\bea
\rho(q) & = & \left(\frac{m\omega}{\pi\hbar}\right)^{\!d/2} 2 \sum_{\mu=0}^M
              F_{M-\mu}^{(d)}(-1)^\mu L_\mu(2x)\, e^{-x}, \label{dden}\\
\xi(q)  & = & \hom\left(\frac{m\omega}{\pi\hbar}\right)^{\!d/2}\frac{d}{2}
              \sum_{\mu=0}^M G_{M-\mu}^{(d)}(-1)^\mu L_\mu(2x)\, e^{-x}.
\label{dkin}
\eea
The coefficients $F_\nu^{(d)}$, $G_\nu^{(d)}$ are given by
\bea
F_\nu^{(d)} & = & \nu + 1 + \sum_{m=1}^{[\nu/2]}
                  (\nu+1-2m)\,g_m^{(d)}\,,\nonumber\\
G_\nu^{(d)} & = & (\nu + 1)^2 + \sum_{m=1}^{[\nu/2]}
                  (\nu+1-2m)^2g_m^{(d)}\,,
\eea
using (for $m\geq1$)
\bea
g_m^{(1)} & = & - (2m-3)!!/(2m)!!\,
                     \quad \hbox{with} \quad g_1^{(1)}=1/2\,,\nonumber\\
g_m^{(2)} & = & 0\,, \qquad g_m^{(3)} = (2m-1)!!/(2m)!!\,, \nonumber\\
g_m^{(4)} & = & 1\,, \qquad g_m^{(5)} = (2m+1)!!/(2m)!!\,,
\eea
and so on. For even dimensions, $F_\nu^{(d)}$ and
$G_\nu^{(d)}$ can be computed analytically, with $[\nu/2] \equiv
{\rm integer}(\nu/2)$.

Like Eqs.\ \eq{den}, \eq{kin}, to which Eqs.\ \eq{dden}, \eq{dkin}
reduce for 2D, the latter are much simpler for numerical
computations than their definitions in terms of the eigenfunctions,
which necessitate multiple summations for $d\ge 2$.

In the TF or local-density approximation (LDA), one gets from the potential
$V(q)=(m\omega^2/2)\,q^2=(\hom/2)\,x$ the following local densities
\bea
\rho_{TF}(q) & = & \frac{4}{d}\,\frac{1}{\Gamma(\frac{d}{2})}
                   \left(\frac{m\omega}{2\pi\hbar}\right)^{\!d/2}
                   (\lambda-x/2)^{d/2},\label{tfden}\\
\tau_{TF}(q) & = & \frac{4\hom}{(d\!+\!2)}\,\frac{1}{\Gamma(\frac{d}{2})}
                   \left(\frac{m\omega}{2\pi\hbar}\right)^{\!d/2}
                   \!(\lambda-x/2)^{d/2+1},\label{tfkin}
\eea
where $\lambda=E_F/(\hom)$. The TF functional relation between $\tau$ and
$\rho$ is then given by
\be
\tau_{TF}[\rho] = \frac{\hbar^2}{2m}\,\frac{4\pi d}{(d+2)}
                  \left[\frac{d}{4}\,\Gamma\!\left(\frac{d}{2}
                  \right)\right]^{\!2/d} \rho^{\,1+2/d}.
\label{tauofrho}
\ee
Our objective now is to study numerically the above relation using the
exact densities and to see how well it holds locally as well as globally
(i.e., upon integration). We have already seen analytically that
its integral yields the exact kinetic energy for $d=2$.

{\it Numerical results. ---} The following figures show numerical results
in units such that $\hbar=\omega=m=1$. We make the following observations:

1. As is well known (cf.\ also Ref.\ \cite{vign}), the densities $\rho(q)$
   and $\tau(q)$, $\tau_1(q)$ oscillate around the smooth TF densities
   \eq{tfden} and \eq{tfkin}, respectively, except near the turning point
   where the latter go to zero (see Figs.\ \ref{1ho}, \ref{2ho}).

2. The shell oscillations in the quantities $\tau(q)$ and $\tau_1(q)$
   are exactly opposite in phase, so that their mean $\xi(q)$ is a smooth
   function of $q$ that (except near the turning point) closely follows
   the TF density $\tau_{TF}(q)$ given in \eq{tfkin} (see \fig{3ho}).
   This has already been observed long ago \cite{rajz}.

3. The functional $\tau_{TF}[\rho(q)]$ in \eq{tauofrho}, using the exact
   densities $\rho(q)$, reproduces locally the exact kinetic energy
   density $\tau(q)$ surprisingly well, {\it including the shell
   oscillations}, except near the classical turning point and in the far
   tail region (see Figs.\ \ref{1ho}, \ref{2ho}).

4. The integral of $\tau_{TF}[\rho(q)]$ over the $d$-dim\-ensional space
   yields the exact kinetic energy only for $d=2$, see Eq.\ \eq{ekinex}.
   In the other cases it yields kinetic energies
   with errors less than one percent for $N \gtrsim$ 14, 100, and
   900 in $d=1$, 3, and 4 dimensions, respectively.

\Figurebb{3ho}{50}{50}{568}{600}{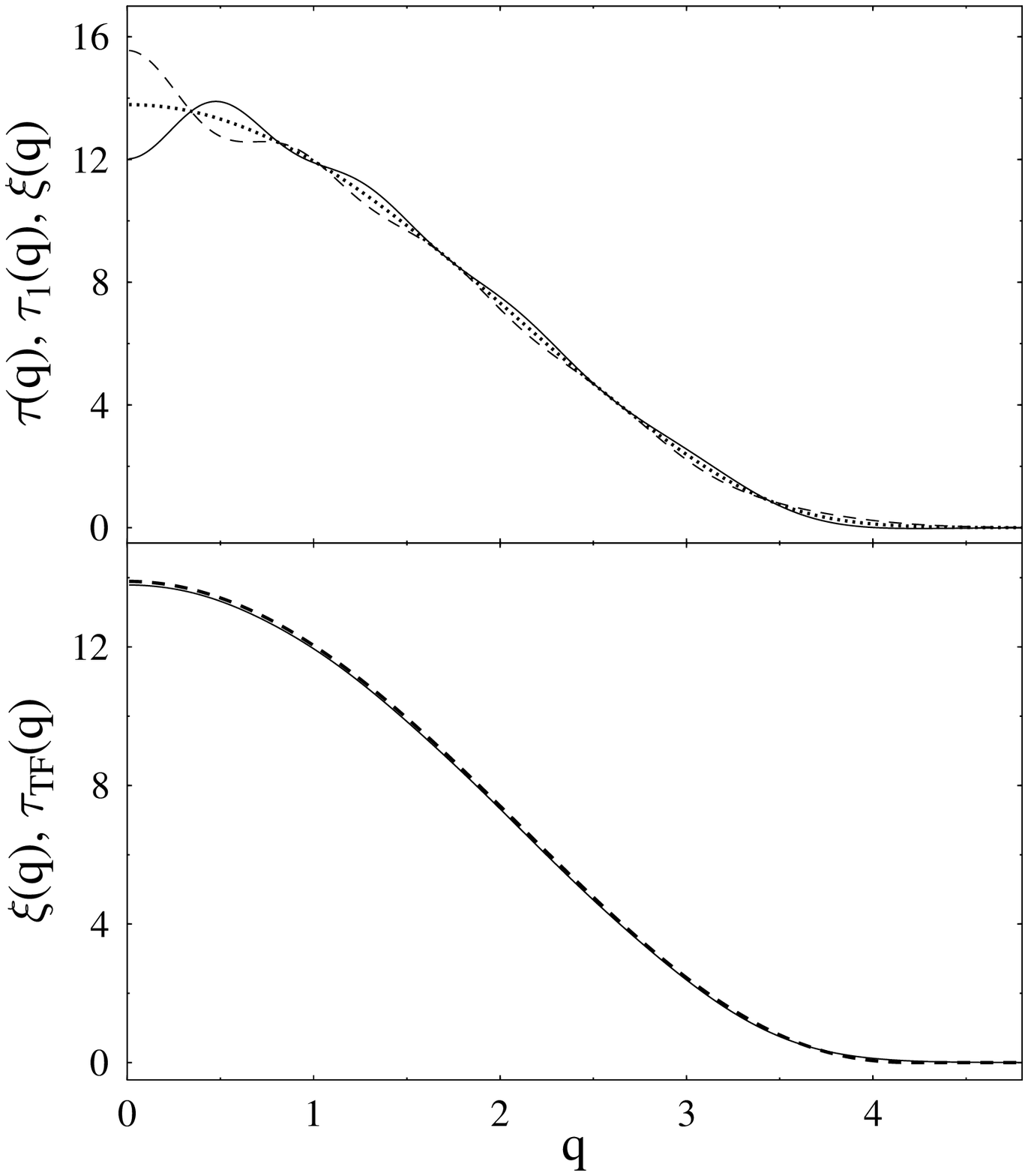}{9}{8.5}{
Kinetic energy densities for $N=240$ particles filling 8 shells of a
3D isotropic harmonic oscillator. {\it Upper panel:}
solid line:
$\tau(q)$, dashed line: $\tau_1(q)$, dotted line: $\xi(q)$. {\it Lower
panel:} solid line: $\xi(q)$, dashed line: $\tau_{TF}(q)$.
}

\Figurebb{1ho}{50}{70}{568}{620}{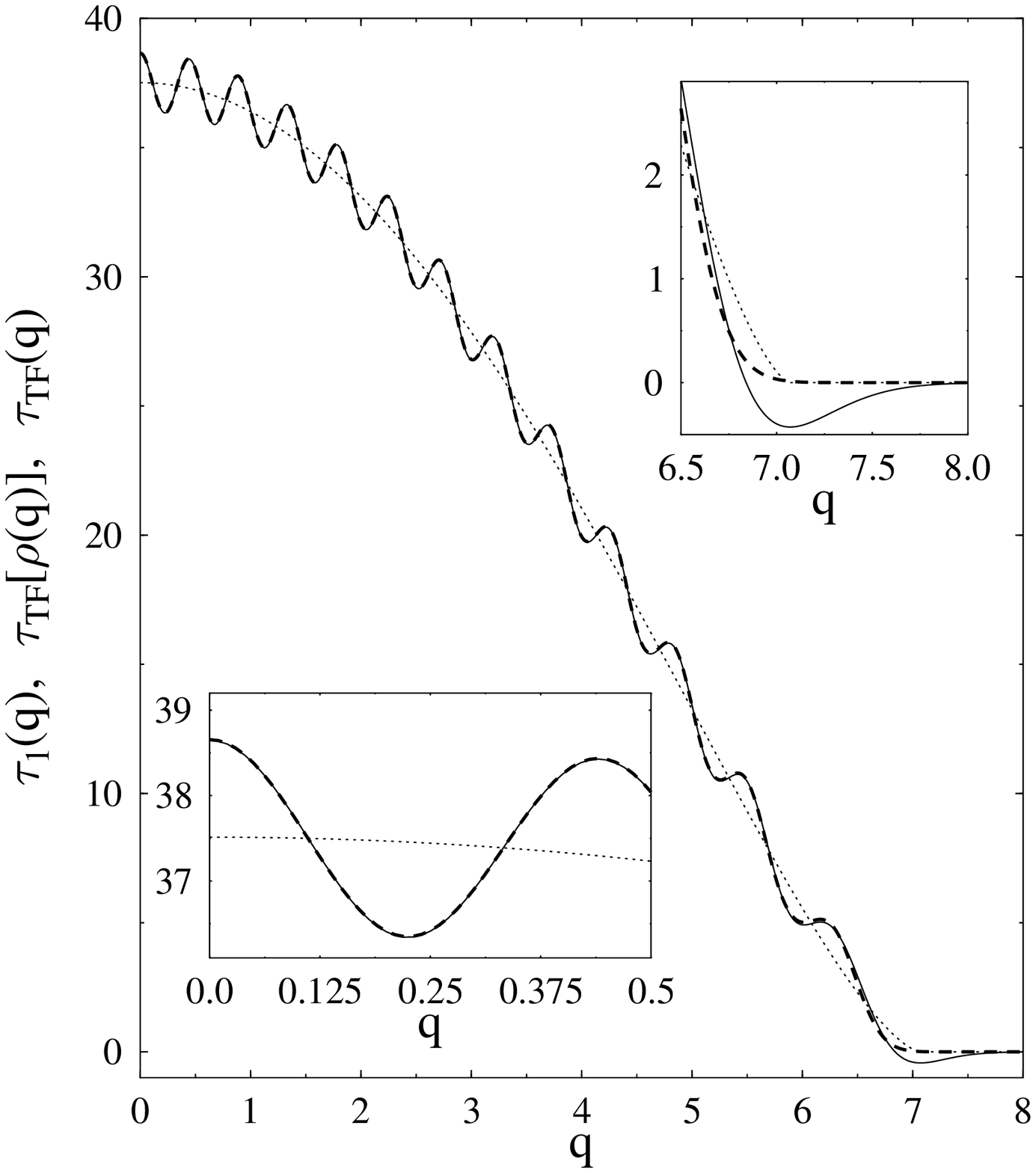}{9.0}{8.5}{
Kinetic energy densities for $N=50$ particles fil\-ling 25 shells of
a 1D harmonic oscillator. Solid lines: exact $\tau(q)$.
Dashed lines: TF relation $\tau_{TF}[\rho(q)]$ \protect\eq{tauofrho}
using the exact $\rho(q)$. Dotted lines: TF density $\tau_{TF}(q)$
\protect\eq{tfkin}. The inserts give close-ups near the center and
the tail region.
}

\vspace{-0.5cm}

\Figurebb{2ho}{50}{70}{568}{620}{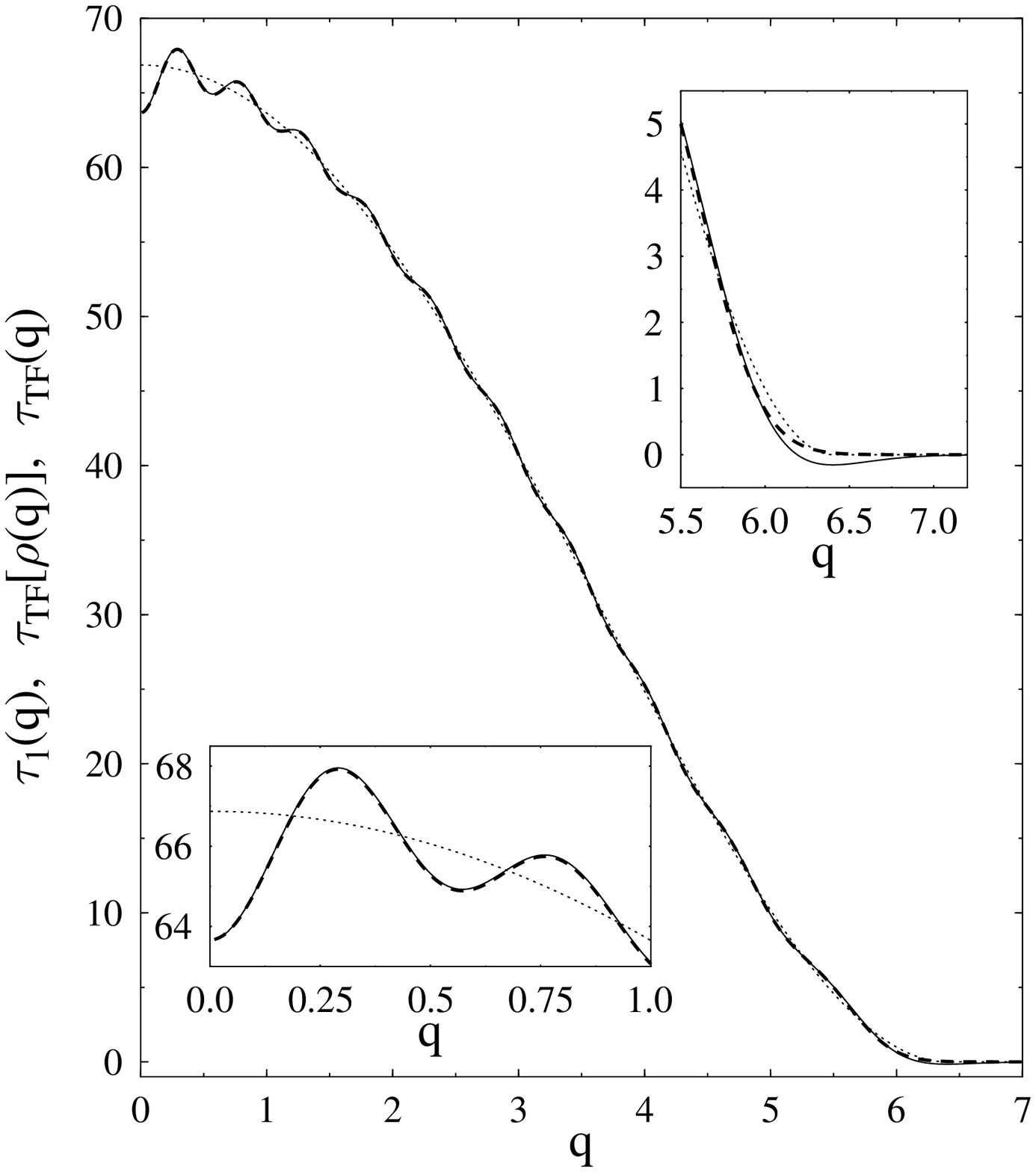}{9}{8.5}{
The same as \protect\fig{1ho}, but for $N=420$ particles fil\-ling 20
shells of a 2D isotropic harmonic oscillator.
}

{\it Summary and conclusions. ---} Our finding that the TF functional
relation $\tau_{TF}[\rho(q)]$ works so well locally is surprising, since
it theoretically is exact only in the LDA,
i.e., for spatially homogeneous systems. That it reproduces the strong
local shell oscillations in $\tau(q)$ so accurately -- in the figures,
the error cannot been recognized except in the tail regions -- is therefore
unexpected and does not seem to have been noticed before \cite{bill}.
We must, however,
add the {\it caveat} that the functional $\tau_{TF}[\rho(q)]$ cannot be
{\it variationally} exact. As is well known, indeed, the Euler-Lagrange
variational equation derived from it leads
precisely to the TF density $\rho_{TF}(q)$ in Eq.\ \eq{tfden}, and
{\it not} to the exact quantum-mechanical density $\rho(q)$.

In the
2D case, where the integral even reproduces the {\it exact} kinetic
energy, our result supports the basic assumptions made in Ref.\
\cite{frac} concerning the inclusion of a short-range two-body
force through fractional statistics, which relies upon the TF relation
\eq{tau2ofrho}.

Finally, we wish to emphasize that
the recent work of Vignolo \etal\ \cite{vign} is a {\it special case}
of our more general results, and point out that the
prominent shell structure displayed in 2D could also become
observable in experiments on alkali vapours.

We are grateful to R. K. Bhaduri and M. V. N. Murthy for encouraging
discussions and acknowledge the warm hospitality of the
Department of Physics and Astronomy at McMaster University.  We would also
like to acknowledge financial support from the
Deutsche For\-schungsgemeinschaft and the NSERC of Canda.

\end{document}